\documentclass[aps,prl,twocolumn,groupedaddress,amsmath,amssymb]{revtex4-1}
\usepackage{mathtools}
\usepackage{graphicx,caption,subcaption}  
\usepackage{dcolumn}   
\usepackage{bm}        
\usepackage{verbatim}   
\usepackage{wrapfig}
\usepackage{float, array}
\usepackage{amsmath}
 \usepackage{amssymb}
\usepackage{verbatim, url}
\usepackage{fullpage}
\usepackage{enumerate}
\usepackage{bm}
\usepackage{notes2bib}
\usepackage{lpic}
\usepackage{pgfplots}
\usepackage[T1]{fontenc}
\usepackage[latin1]{inputenc}
\usepackage[extra]{tipa}
\usepackage{accents}
\usepackage{epstopdf}

 \usepackage{amsthm}

\newlength{\dhatheight}

\newcommand{\nn}{\nonumber}
\newcommand{\f}[2] {\frac{#1}{#2}}
\newcommand{\p}{\partial}

\newcommand{\beq}{\begin{equation}}
\newcommand{\eeq}{\end{equation}}
\newcommand{\beqn}{\begin{eqnarray}}
\newcommand{\eeqn}{\end{eqnarray}}

\newcommand{\tl}{\tilde}

\newcommand{\te}{\tilde{\epsilon}}

\DeclarePairedDelimiter\abs{\lvert}{\rvert}%
\DeclarePairedDelimiter\norm{\lVert}{\rVert}%

\makeatletter
\let\oldabs\abs
\def\abs{\@ifstar{\oldabs}{\oldabs*}}
\let\oldnorm\norm
\def\norm{\@ifstar{\oldnorm}{\oldnorm*}}
\makeatother

\begin{document}

\title{Tunnel Splitting in Asymmetric Double Well Potentials : An Improved WKB Calculation}
\author{Seyyed M.H. Halataei, Anthony J. Leggett}
\affiliation{
   Department of Physics, University of Illinois at Urbana-Champaign,\\
   1110 West Green St, Urbana, Illinois 61801, USA
   }
   
\date{Oct 24, 2017}
\begin{abstract}
We present an improved Wentzel-Kramers-Brillouin (WKB) calculation of tunnel splitting in one dimensional asymmetric double well potentials. We show the tunnel splitting in general can have linear dependence to bias energy beside the well-known quadratic dependence. We demonstrate that the linear correction is greater than previously thought.  
 \end{abstract}

\maketitle

\section{INTRODUCTION}
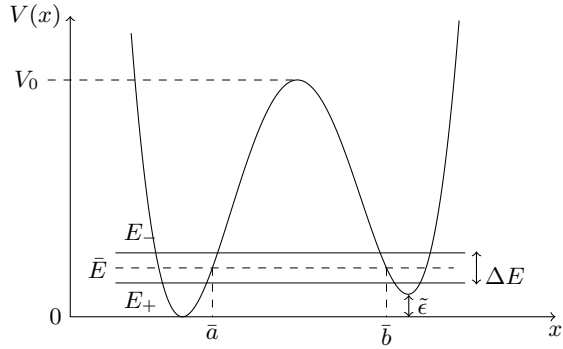
\begin{figure}[b]
\centering
\begin{tikzpicture}[xscale=1.5]
\draw [<->] (-2.7+.7,4) node[left] {$V(x)$} -- (-2.7+.7,0) node[left] {$0$} -- (2.7-.4,0) node[below] {$x$};
\draw [domain=-1.46:1.45, samples=300] plot (\x,{3* (\x*\x-1)*(\x*\x-1) + .3/2*(\x+1)});
\draw [ultra thin,dashed](-3+.8,3+.3/2) node[left] {$V_0$} -- (0,3+.3/2);
\draw [very thin,black] (-2+.4,1/2+.3/2-.2) node [below right] {$E_{+}$} -- (1.9-.4,1/2+.3/2-.2) ;
\draw [very thin,black] (-2+.4,1/2+.3/2+.2)  node [above right] {$E_{-}$} -- (1.9-.4,1/2+.3/2+.2);
\draw [ultra thin, dashed] (-1.8+.2,1/2+.3/2)  node [left] {$\bar{E}$} -- (1.8-.4,1/2+.3/2);
\draw [<->] (+1.8+.2-.4,1/2+.3/2+.22) -- (+1.8+.2-.4,1/2+.3/2) -- node [right] {$\Delta E$} (+1.8+.2-.4,1/2+.3/2-.22) ;
\draw [<->] (1,0) -- (1,0.3/2-.012/2) node[right] {$\tilde{\epsilon}$} -- (1,0.3-.012);
\draw [ultra thin,dashed] (-.74,1/2+.3/2) -- (-.74,0) node[below] {$\bar{a}$} ;
\draw [ultra thin, dashed] (.8026,1/2+.3/2) -- (0.8026,0) node[below] {$\bar{b}$} ;

\end{tikzpicture}
\caption{Asymmetric double well potential $V(x)$ considered in this paper. $E_\pm$ are energy levels of the ground state doublet. $\Delta E$ is the level splitting and $\tilde{\epsilon}$ is the bias in the bottoms of the wells. $\bar{E}$ is a mean energy between $E_-$ and $E_+$ which is used as a mathematical tool to calculate $\Delta E$. $\bar{a}$ and $\bar{b}$ are turning points for the fictitious energy level $\bar{E}$. The height of barrier is $V_0$ which is much larger than other energy quantities in the problem.}
\label{Fig1}
\end{figure}

\begin{figure}[b]
\centering
\begin{tikzpicture}[xscale=1.5]

\draw [<->] (-2.5+.5,4) node[left] {$V(x)$} -- (-2.5+.5,0) node[left] {$0$} -- (2.5-.7,0) node[below] {$x$};
\draw [domain=-1.46:1.45, samples=300] plot (\x,{3* (\x*\x-1)*(\x*\x-1) + .3/2*(\x+1)});
\draw [dashed] (-1.19,1/2) -- (-.778,1/2) ;
\draw [<->] (-1.19-.1,0) -- (-1.19-.1,1/4) node[left] {$\f{\hbar \omega_L}{2}$} -- (-1.19-.1,1/2);

\draw [dashed] (0.7598,1/2+.3) -- (1.18193,1/2+.3);
\draw [<->] (1.18193+.2,.3) -- (1.18193+.2,.3+1/4) node[right] {$\f{\hbar \omega_R}{2}$} -- (1.18193+.2,1/2+.3);

\draw [<->] (0,1/2) -- (0,1/2+.3/2) node[right] {$\epsilon$} -- (0,1/2+.3);

\draw [<->] (1,0) -- (1,0.3/2-.012/2) node[right] {$\tilde{\epsilon}$} -- (1,0.3-.012);

\end{tikzpicture}
\caption{Asymmetric double well potential $V(x)$. $\omega_L$ ($\omega_R$) is the small oscillation frequencies in the left (right) well. In the absence of tunneling, $\hbar \omega_L/2$ ($\hbar \omega_R / 2 + \tilde{\epsilon}$) is the ground-state energy of the state localized in the left (right) well and $\epsilon$ is the difference between these two energies. }
\label{Fig2}
\end{figure}
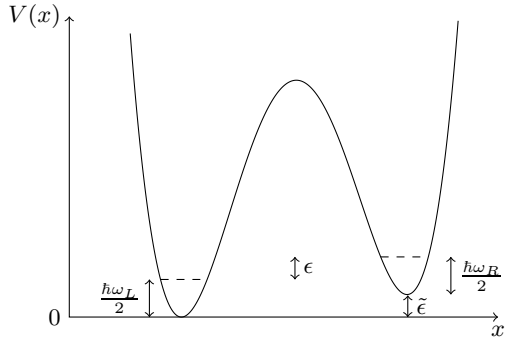

The purpose of this paper is to calculate the energy level splitting (or tunnel splitting), $\Delta E$, in a smooth, asymmetric, one-dimensional potential, such as that in Fig. \ref{Fig1}, to first order in $\te/\hbar \omega$ where $\te$ is the bias energy between the bottom of the wells and $\omega$ is the order of magnitude of the small oscillation frequencies $\omega_R$, $\omega_L$ in the right and the left wells (see Figs. \ref{Fig1}-\ref{Fig2}). 

The problem of quantum mechanical tunneling in a double-well potential is ubiquitous in physics. The quantum state of the system in such problems is effectively restricted to a two-dimensional Hilbert space. Quantum tunneling allows the state to hop between these two dimensions. Apart from the well-known microscopic example of inversion of an amonia molecule, in recent decades quantum tunneling has been observed in macroscopic phenomena such as the tunneling of magnetic flux in an rf SQUID\cite{Leggett87, vanderWal2000, Friedman2000, Clarke2008}, tunneling of Bose-Einstein condensates \cite{Schumm2005, Hall2007} and electronic spin tunneling in the nano-magnetic molecules such as Fe8 \cite{ Wernsdorfer1999, Gatteschi2006, Takahashi2011}. 

In some problems the height of the barrier $V_0$ is much larger than the energy gap $\hbar \omega$ between the ground state doublet and higher excited states. WKB approximation can be applied \emph{under} the barrier in these problems. However, application of WKB \emph{inside} the wells gives inaccurate results \cite{Garg2000}. The reason is that, crudely speaking, the semi classical approximation of WKB is suitable where the classical momentum of a particle $\abs{p(x)} = \sqrt{2 m \abs{E - V(x)}}$ is large. This is not satisfied for a particle in ground state inside a well. However, under the barrier since $V(x)$ is large the condition is satisfied and one can employ the WKB approximation. \cite{Garg2000} 

Previous works \cite{Dekker87,Garg2001,Song2008,Song2015} have calculated the energy splitting $\Delta E$ and tunneling amplitude $\Delta$ in an asymmetric potential to zeroth order in $\epsilon/\hbar \omega$ and $\epsilon/V_0$. The general belief \cite{Dekker87} is that the correction to these quantities are of order $\epsilon / V_0$. It is also implicitly assumed in the bulk of literature that the tunneling amplitude $\Delta$ is relatively independent of the bias energy $\tl{\epsilon}$ or $\epsilon$.

In the present paper, however, we show that the correction to tunnel splitting is in general of order $\tl{\epsilon}/\hbar \omega$ in the WKB limit, rather than $\tl{\epsilon}/V_0$.

Calculations of this paper are more accurate than its previous counterparts. For example, we give an expression for  $\Delta$, Eq. (\ref{Delta_asym}) ,which does not depend on the value of $\Delta$ itself. The situation is rather different in Ref. \cite{Garg2001,Song2008, Song2015}. The Gamow factor, $e^{-2 I}$, in those references depends on the actual energy of the levels $E_\pm$ and, hence, on the value of the $\Delta E$, and $\Delta$. Ref. \cite{Garg2001} discusses that this dependence is rather weak. Here, however, we obtain $\Delta$ as a function of the Gamow factor of a fictitious energy $\bar{E}$, Eq. \eqref{Ebar0}, independent of $\Delta E$, $\Delta$. Furthermore we show quantitatively that the correction to our expression is negligible. 

Before embarking on detailed calculations in the following sections, let us summarize the main results of this paper for the energy splitting and tunneling amplitude of ground state doublet. We denote the energy of the near even parity state in the doublet by $E_+$, the lower level, and the energy of the nearly odd parity state, the upper level by $E_-$ (Fig. \ref{Fig1}). Then we shall derive that the energy splitting between these two levels $\Delta E = E_- - E_+$ is 
\beq
\Delta E = \sqrt{\epsilon^2 + \Delta^2} \ ,
\eeq
where
\beq \label{epsilon}
\epsilon = \tilde{\epsilon} + \f{\hbar (\omega_R - \omega_L)}{2},
\eeq
$\tilde{\epsilon}$ is the energy difference between the minima of the potential, $\epsilon$ is the energy difference between the ground states of the particle in each well in absent of tunneling (Fig. \ref{Fig1}-\ref{Fig2}) and 
\beq
\label{Delta_asym}
\Delta = \f{\hbar \sqrt{\omega_R \omega_L}}{\sqrt{e \pi}} (1 + \f{k}{4}\f{\epsilon}{\hbar \omega_L} \f{\omega_R - \omega_L}{\omega_R})e^{-\bar{I}(\bar{E}(\te))}  ,
\eeq
where
\beqn
\label{k} & &k = \gamma - \ln 2 \simeq -0.11, \\ \label{Ebar0}
& &\bar{E}(\te) = \bar{E} =  \f{\hbar (\omega_L + \omega_R)}{4} + \f{\tilde{\epsilon}}{2}, \\
& &I(\bar{E}(\te)) = \f{1}{\hbar} \int\limits_{
\bar{a}}^{\bar{b}} |p| dx. 
\eeqn
$\gamma$ is the Euler-Mascheroni constant and 
\beq
p = \sqrt{2 m ( \bar{E} - V(x))}.
\eeq
Also, $\bar{a}$ and $\bar{b}$ are the turning points for a classical particle with energy $\bar{E}$ which wishes to climb up the barrier from either well (See Fig. \ref{Fig1}). 

\section{Wave functions and Energy Quantization Equations} \label{Wf}
\begin{figure}[h]
\centering
\begin{tikzpicture}[xscale=2.3]

\draw [<->] (-1.5,4) node[left] {$V(x)$} -- (-1.5,0) node[left] {$0$} -- (1.5,0) node[below] {$x$};
\draw [domain=-1.4:1.4, samples=300] plot (\x,{6* (\x*\x-1)*(\x*\x-1) + .1/2*(\x+1)});
\draw [dashdotted] (-1.00104,0) -- (-1.00104,-.5) node [below] {$x_L$};
\draw [dashdotted] (0.998957,.1) -- (0.998957,-.5) node [below] {$x_R$};


\draw [densely dotted] (.7,0) -- (.7,4);
\draw [dashed] (.6,0) -- (.6,4);
\draw [dashed] (1.3,0) -- (1.3, 4);
\draw [<->] (.6,3) -- (.95,3) node [above] {$R$} -- (1.3,3);

\draw [<->] (-.65,1) -- (0,1) node [above] {$B$} -- (.7,1);

\draw [densely dotted] (-.65,0) -- (-.65,4);
\draw [dashed] (-.55,0) -- (-.55,4);
\draw [dashed] (-1.3,0) -- (-1.3, 4);

\draw [<->] (-1.3,3) -- (-.925,3) node [above] {$L$} -- (-.55,3);

\draw [<-> ](-.55,4.1) -- (-.6,4.1) node [above] {$LB$} -- (-.65,4.1)  ;

\draw [<-> ](.6,4.1) -- (.65,4.1) node [above] {$BR$} -- (.7,4.1)  ;

\draw [dashdotted] (-1.5,1/2) node [left] {$E_+$} -- (1.5,1/2) ;
\draw [dashdotted] (-.844734,1/2-.1/2) -- (-.844734,0) node [below] {$a_+$};

\draw [dashdotted] (0.859972,1/2-.1/2) -- (0.859972,0) node [below] {$b_+$};



\end{tikzpicture}
\caption{Regions \textit{L} and \textit{R} denote the domain of potential near the left and right minima in which the potential is sufficiently quadratic. Region \textit{B} is under the barrier area away from the turning points in which the WKB approximation can be applied. \textit{LB} and \textit{BR} are the overlapping areas where we match the wave function of each region to that of its neighboring region. $x_L$ and $x_R$ denote the coordinates of the local minima. }
\label{Fig3}
\end{figure}
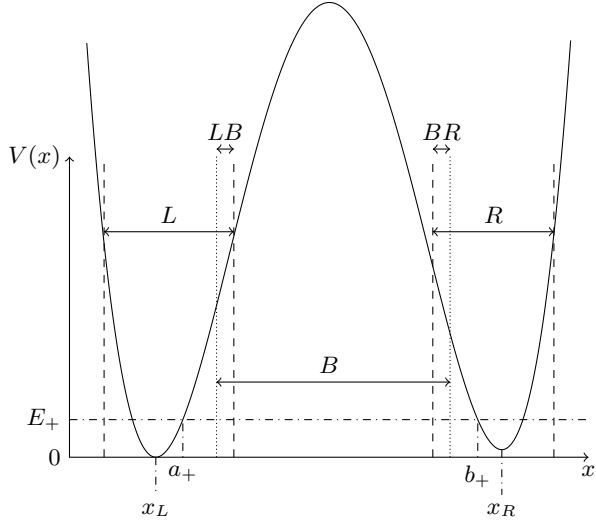

In order to find the energy level splitting in the ground state doublet we find the wave function near the left minimum in region \textit{L}, under the barrier in region \textit{B}, and near the right minimum in region \textit{R} as illustrated in Fig. \ref{Fig3}. Then we connect these wave functions in the overlapping regions \textit{LB} and \textit{BR}. The connection formulas give us a constraint which determines the energy splitting. 

We assume that the potential is nearly parabolic near the minima in the regions \textit{L} and \textit{R}. The Schrodinger equation for parabolic potentials can be solved exactly for any given energy. The solutions are parabolic cylinder functions. We find these solutions such that they do not diverge as $x \rightarrow \pm \infty$, to avoid violation of square-integrability of the wave function. 

Under the barrier, in region \textit{B}, we use the WKB approximation method.

\subsection{Wave functions near the local minima of potential: Parabolic cylinder functions}
As discussed earlier, near the minima of the potential $x_L$ and $x_R$ we can write 
\beq
\label{V}
V(x) =  \begin{cases} 
      \f{1}{2} m \omega_L^2 (x-x_L)^2 + \dotsm & x \in L \\
      \tilde{\epsilon} + \f{1}{2} m \omega_R^2 (x-x_R)^2 + \dotsm & x \in R  
   \end{cases} 
\eeq
where 
\beqn
\omega_L^2 &=& \f{V''(x_L)}{m}, \\
\omega_R^2 &=& \f{V''(x_R)}{m},
\eeqn
and the zero-point of potential is set such that $V(x_L)=0$ and $V(x_R) = \tilde{\epsilon}$ (Fig. \ref{Fig3}). 

Neglecting the higher order terms in the potential, the Schrodinger equation in region \textit{L} for the lower level in the ground state doublet with energy $E_+$ becomes
\beq
\label{schrodinger}
-\f{\hbar^2}{2m} \f{d^2}{dx^2}\psi_L(x) + \f{1}{2} m \omega_L^2 (x-x_L)^2 \psi_L (x) = E_+ \psi_L (x).
\eeq
We can write the above equation in the form of differential equation of parabolic cylinder functions by defining,
\beqn
\label{eta_l} \eta_{\scriptscriptstyle l} &\equiv& \f{x - x_L}{\sqrt{\hbar/2 m \omega_L}}, \\ \label{p0}
\zeta_+^{\scriptscriptstyle L} &\equiv& \f{E_+}{\hbar \omega_L} - \f{1}{2}.
\eeqn
Notice that $\eta_{\scriptscriptstyle l}$ is a variable and varies with $x$ while $x_L$ is a fixed point (We use lowercase l (r) for variables of the left (right) well and uppercase L (R) for its fixed quantities.). With the above definitions, Eq. \eqref{schrodinger} can be rewritten as 
\beq
\f{d^2}{d\eta_{\scriptscriptstyle l}^2}\psi_L(\eta_{\scriptscriptstyle l}) + \left(\zeta_+^{\scriptscriptstyle L} + \f{1}{2} - \f{\eta_{\scriptscriptstyle l}^2}{4} \right) \psi_L (\eta_{\scriptscriptstyle l}) = 0.
\eeq
which is manifestly parabolic cylinder functions' differential equation \cite{Gradshteyn2000}. This  equation has two independent solutions, $D_{\zeta_+^{\scriptscriptstyle L}}(\eta_{\scriptscriptstyle l})$ and $D_{\zeta_+^{\scriptscriptstyle L}}(-\eta_{\scriptscriptstyle l})$. The former diverges as $\eta_{\scriptscriptstyle l}\rightarrow -\infty$ and is not allowed by square-integribility condition. Therefore, the physical solution for the wave function in region $L$ is
\beq
\psi_L (\eta_{\scriptscriptstyle l}) = \alpha_L D_{\zeta_+^{\scriptscriptstyle L}} (-\eta_{\scriptscriptstyle l})
\eeq
where $\alpha_L$ is a constant to be determined by matching conditions below. 

For real values of $\eta_{\scriptscriptstyle l}$ the asymptotic expansion of $D_{\zeta_+^{\scriptscriptstyle L}} (-\eta_{\scriptscriptstyle l})$ to leading order, when $|\zeta_+^{\scriptscriptstyle L}| \ll 1$, is \cite{Gradshteyn2000}
\begin{align}
\nn & D_{\zeta_+^{\scriptscriptstyle L}} (-\eta_{\scriptscriptstyle l})= \\  &\begin{cases} 
      (-1)^{\zeta_+^{\scriptscriptstyle L}} \  \eta_{\scriptscriptstyle l}^{\zeta_+^{\scriptscriptstyle L}} e^{-\eta_{\scriptscriptstyle l}^2/4}, & \eta_{\scriptscriptstyle l} \!\ll\! -1 \\
      \cos (\pi \zeta_+^{\scriptscriptstyle L}) |\eta_{\scriptscriptstyle l}|^{\zeta_+^{\scriptscriptstyle L}} e^{-\eta_{\scriptscriptstyle l}^2/4}  + \f{\sqrt{2 \pi}}{\Gamma(-\zeta_+^{\scriptscriptstyle L})} \f{e^{\eta_{\scriptscriptstyle l}^2/4}}{|\eta_{\scriptscriptstyle l}|^{\zeta_+^{\scriptscriptstyle L} + 1}}  & \eta_{\scriptscriptstyle l} \!\gg\! 1  
   \end{cases} 
\end{align}
which exponentially decays as $\eta_{\scriptscriptstyle l}\rightarrow -\infty$ and has a decaying and growing parts for positive large $\eta_{\scriptscriptstyle l}$. We shall see shortly that keeping both of these parts is necessary for matching the wave functions under the barrier and near the right well.

The quantity $|\zeta_+^{\scriptscriptstyle L}|$ is much smaller than unity. One can observe this fact by noting that in the unbiased symmetric case this quantity is half of the tunneling amplitude $\Delta$ which is exponentially small. In the asymmetric case definition \eqref{p0} implies that $|\zeta_+^{\scriptscriptstyle L}|$ is at most of order $\tilde{\epsilon}/\hbar \omega$ which is much smaller than unity by our convention in this paper (We demonstrate this fact rather more rigorously in Sec. \ref{EnergySplitting} when we find the energy levels).  

Similarly we can find the wave function near the right well in region \textit{R}. The Schrodinger equation to second order approximation of the potential is 
\begin{align}
\nn -\f{\hbar^2}{2m} \f{d^2}{dx^2}\psi_R(x) + \left(\f{1}{2} m \omega_L^2 (x-x_R)^2 + \tilde{\epsilon}\right) \psi_R (x) \\ \label{schrodingerR}
 = E_+ \psi_R (x).
\end{align}
By defining,
\beqn
\label{eta_r} \eta_{\scriptscriptstyle r} &\equiv& \f{x - x_R}{\sqrt{\hbar/2 m \omega_R}}, \\
\label{q0} \zeta_+^{\scriptscriptstyle R} &\equiv& \f{E_+ - \tilde{\epsilon}}{\hbar \omega_R} - \f{1}{2},
\eeqn
Eq. \eqref{schrodingerR} can be written as 
\beq
\label{diffR}
\f{d^2}{d \eta_{\scriptscriptstyle r}^2}\psi_R(\eta_{\scriptscriptstyle r}) + \left(\zeta_+^{\scriptscriptstyle R} + \f{1}{2} - \f{\eta_{\scriptscriptstyle r}^2}{4} \right) \psi_R (\eta_{\scriptscriptstyle r}) = 0,
\eeq
which has two solutions $D_{\zeta_+^{\scriptscriptstyle R}}(\eta_{\scriptscriptstyle r})$ and $D_{\zeta_+^{\scriptscriptstyle R}}(-\eta_{\scriptscriptstyle r})$. This time we reject the latter as it diverges when $\eta_{\scriptscriptstyle r} \rightarrow \infty$. So the physical solution of \eqref{diffR} is
\beq
\psi_R (\eta_{\scriptscriptstyle r}) = \alpha_R D_{\zeta_+^{\scriptscriptstyle R}} (\eta_{\scriptscriptstyle r})
\eeq
where $\alpha_R$ is a coefficient to be determined and the asymptotic expansion of $D_{\zeta_+^{\scriptscriptstyle R}} (\eta_{\scriptscriptstyle r})$ for $|\zeta_+^{\scriptscriptstyle R}| \ll 1$ is 
\begin{align}
\nn & D_{\zeta_+^{\scriptscriptstyle R}} (\eta_{\scriptscriptstyle r}) =\\ & \begin{cases} 
       \cos (\pi \zeta_+^{\scriptscriptstyle R}) |\eta_{\scriptscriptstyle r}|^{\zeta_+^{\scriptscriptstyle R}} e^{-\eta_{\scriptscriptstyle r}^2/4} + \f{\sqrt{2 \pi}}{\Gamma(-\zeta_+^{\scriptscriptstyle R})} \f{e^{\eta_{\scriptscriptstyle r}^2/4}}{|\eta_{\scriptscriptstyle r}|^{\zeta_+^{\scriptscriptstyle R} + 1}}  & \eta_{\scriptscriptstyle r} \ll -1  \\
      (-1)^{\zeta_+^{\scriptscriptstyle R}} \  \eta_{\scriptscriptstyle r}^{\zeta_+^{\scriptscriptstyle R}} e^{-\eta_{\scriptscriptstyle r}^2/4} & \eta_{\scriptscriptstyle r} \gg 1 
   \end{cases} 
\end{align}
This again decays nicely for large positive $\eta_{\scriptscriptstyle r}$ and has growing and decaying components for large negative $\eta_{\scriptscriptstyle r}$ as expected. $|\zeta_+^{\scriptscriptstyle R}|$ is also much smaller than unity for the problem we consider for the same reasons mentioned above for $|\zeta_+^{\scriptscriptstyle L}|$.

The particular regions of interest are $LB$ and $BR$ (Fig. \ref{Fig3}). We use these regions to match $\psi_L$ and $\psi_R$ to the WKB solution under the barrier. These regions are reasonably far from the turning points to satisfy validity condition of the WKB approximation, but yet close enough to the bottom of the wells to allow parabolic approximation of the potential to be employed. In these regions the wave functions that we found in this section is as follows 
\begin{align}
\nn & \psi_+^{par} (x) = \\ \label{psi0} & \begin{cases} 
		\alpha_L \f{\cos (\pi \zeta_+^{\scriptscriptstyle L}) |\eta_{\scriptscriptstyle l}|^{\zeta_+^{\scriptscriptstyle L}}} {e^{\eta_{\scriptscriptstyle l}^2/4}} + \alpha_L \f{\sqrt{2 \pi}}{\Gamma(-\zeta_+^{\scriptscriptstyle L})} \f{e^{\eta_{\scriptscriptstyle l}^2/4}}{|\eta_{\scriptscriptstyle l}|^{\zeta_+^{\scriptscriptstyle L} + 1}}&\! x \in LB  \\
      \alpha_R \f{\cos (\pi \zeta_+^{\scriptscriptstyle R}) |\eta_{\scriptscriptstyle r}|^{\zeta_+^{\scriptscriptstyle R}}}{e^{\eta_{\scriptscriptstyle r}^2/4}} + \alpha_R \f{\sqrt{2 \pi}}{\Gamma(-\zeta_+^{\scriptscriptstyle R})} \f{e^{\eta_{\scriptscriptstyle r}^2/4}}{|\eta_{\scriptscriptstyle r}|^{\zeta_+^{\scriptscriptstyle R} + 1}}  & x \in BR 
   \end{cases} 
\end{align}
where superscript \textit{par} is for \textit{parabolic cylinder} and the relation between $x$ and $\eta_{\scriptscriptstyle l}$, $\eta_{\scriptscriptstyle r}$ is given in Eqs. (\ref{eta_l}) and (\ref{eta_r}).
In the next subsection we shall find WKB wave function under the barrier and match it with Eq. \eqref{psi0}. 
\subsection{Wave functions under the barrier: WKB approximation limit}
Under the barrier, far enough from the turning points in region $B$, we can apply the WKB approximation. With the usual ansatz of $\psi = exp(i \sigma / \hbar)$ one obtains \cite{Landau58} 
\beq
\label{WKB}
\psi^{\text{WKB}}_+(x) = \f{C}{\sqrt{|v(x)|}} e^{-\int\limits_{a}^{x} |p| dx /\hbar} + \f{C'}{\sqrt{|v(x)|}} e^{+\int\limits_{b}^{x} |p| dx /\hbar}
\eeq

where $C$, $C'$ are constants,
\beqn
p(x) &=& \sqrt{2 m (E_+ - V(x))} \ , \\
v(x) &=& \f{p(x)}{m},
\eeqn
and $a_+$,$b_+$ are classical turning points for energy $E_+$ as shown in Fig. \ref{Fig3}. The choice of lower bounds of the integrals in \eqref{WKB} is arbitrary. We chose to use $a_+$,$b_+$ to simplify future equations. This choice is different from what previous authors have used \cite{Garg2000,Song2008,Song2015}. 

The particular regions of interest are again $LB$ and $BR$. In these regions we can approximate the potential with parabolic functions of Eq. \eqref{V}. Under this approximation, e.g., 
\beq \label{E0}
E_+= V(a_+) \simeq \f{1}{2} m \omega_L^2 (a_+ - x_L)^2
\eeq
and one obtains in region $LB$ for the classical momentum
\beq \label{p}
\abs{p(x)}  \simeq  m \omega_L \sqrt{(x - x_L)^2 - (a_+ - x_L)^2}
\eeq
The first integral in \eqref{WKB} can now be taken for $x \in \textit{LB}$. We follow methods developed in Ref. \cite{Garg2000} in taking this integral. The result is 
\beqn
\nn \f{1}{m \omega_L} \int\limits_{a_+}^{x} \abs{p} dx &\simeq& \f{1}{2} (x-x_L)^2 - \f{1}{4} (a_+-x_L)^2 \\ 
\nn &-& \f{1}{2} (a_+ - x_L)^2 \ln \left(\f{2 (x-x_L)}{a_+ - x_L} \right). \\ \label{int}
\eeqn
We used the fact that in the region { \it LB}, $(x-x_L) \gg (a_+ - x_L)$. Nevertheless we kept the second term in Eq. \eqref{int} as the left hand side integral appears in the \textit{exponent} of the first term in Eq. \eqref{WKB}. However, one does not need to keep the similar term in calculating $v(x)$ from Eq. \eqref{p}, as $v(x)$ appears in the denominators in Eq. \eqref{WKB} (not in the exponents \cite{Garg2000}),
\beq
v(x) \simeq \omega_L (x-x_L).
\eeq
For the second integral in Eq. \eqref{WKB} and $x \in \textit{LB}$ we note that 
\beq \label{pbint}
\int\limits_{b_+}^{x} \abs{p} dx = -\int\limits_{a_+}^{b_+} \abs{p} dx + \int\limits_{a_+}^{x} \abs{p} dx.
\eeq
We define,
\beqn
I_+ &\equiv& \int\limits_{a_+}^{b_+} \abs{p} dx, \\
g(\zeta) &\equiv& \sqrt{2 \pi} \left( \zeta + \f{1}{2}\right)^{ \zeta + \f{1}{2}} e^{- (\zeta + \f{1}{2})}.
\eeqn
Now by substituting from Eqs. (\ref{int}-\ref{pbint}) into Eq. \eqref{WKB} and using Eqs. (\ref{eta_l}-\ref{p0}) and (\ref{E0}) we obtain for $x \in \textit{LB}$,
\beq \label{WKBLB}
\psi_+^{\text{WKB}} (x) \simeq K_L \eta_{\scriptscriptstyle l}^{\zeta_+^{\scriptscriptstyle L}} e^{-\eta_{\scriptscriptstyle l}^2/4} + K'_L \eta_{\scriptscriptstyle l}^{-(\zeta_+^{\scriptscriptstyle L}+1)} e^{-\eta_{\scriptscriptstyle l}^2/4},
\eeq
where
\beqn
K_L &=& (\f{\hbar \omega_L}{\pi m})^{-\f{1}{4}} (\f{g_+^{\scriptscriptstyle L}}{2})^{-\f{1}{2}} \  C, \\
K'_L &=& (\f{\hbar \omega_L}{\pi m})^{-\f{1}{4}} (\f{g_+^{\scriptscriptstyle L}}{\pi})^{\f{1}{2}}  e^{-I_+} \ C',
\eeqn
and $g_+^{\scriptscriptstyle L} = g(\zeta_+^{\scriptscriptstyle L})$. 

Similar procedure can be used for region $BR$ under the barrier and near the right well. Most of the equations transform trivially if we make the substitution $E_+ \rightarrow E_+ - \tilde{\epsilon}$. For the energy and momentum one has
\beqn \label{E0R}
E_+ - \tilde{\epsilon}= V(b_+) - \tilde{\epsilon} \simeq \f{1}{2} m \omega_R^2 (x_R - b_+)^2 , \\
\abs{p(x)}  \simeq  m \omega_R \sqrt{(x - x_R)^2 - (x_R - b_+)^2}.
\eeqn  
These can be used to take the second integral in Eq. (\ref{WKB}) as follows
\beqn
\nn \f{1}{m \omega_R} \int\limits_{b_+}^{x} \abs{p} dx &\simeq& - \f{1}{2} (x-x_R)^2 + \f{1}{4} (x_R - b_+)^2 \\ 
\nn &+& \f{1}{2} (x_R - b_+)^2 \ln \left(\f{2 (x_R-x)}{x_R - b_+} \right). \\ \label{intR}
\eeqn
Note that all the signs in the right hand side are flipped in comparison to Eq. \eqref{int}. In taking the integrals \eqref{int} and \eqref{intR} one may use Eq. (2.27) of Ref. \cite{Gradshteyn2000}. In \textit{BR} the velocity is approximately 
\beq
v(x) \simeq \omega_R (x_R - x)
\eeq
and the first integral in Eq. \eqref{WKB} for $x \in \textit{BR}$ can be calculated by the identity
\beq \label{pbintR}
\int\limits_{a_+}^{x} \abs{p} dx = \int\limits_{a_+}^{b_+} \abs{p} dx + \int\limits_{b_+}^{x} \abs{p} dx.
\eeq
Now by substituting from Eqs. (\ref{intR}-\ref{pbintR}) into Eq. \eqref{WKB} and using Eqs. (\ref{eta_r}-\ref{q0}) and (\ref{E0R}) we obtain for $x \in \textit{BR}$,
\beq \label{WKBBR}
\psi_+^{\text{WKB}} (x) \simeq L_R \abs{\eta_{\scriptscriptstyle r}}^{\zeta_+^{\scriptscriptstyle R}} e^{-\eta_{\scriptscriptstyle r}^2/4} + L'_R \abs{\eta_{\scriptscriptstyle r}}^{-(\zeta_+^{\scriptscriptstyle R}+1)} e^{-\eta_{\scriptscriptstyle r}^2/4},
\eeq
where
\beqn
L_R &=& (\f{\hbar \omega_R}{\pi m})^{-\f{1}{4}} (\f{g_+^{\scriptscriptstyle R}}{\pi})^{\f{1}{2}}  e^{-I_+} C, \\ 
L'_R &=& (\f{\hbar \omega_R}{\pi m})^{-\f{1}{4}} (\f{g_+^{\scriptscriptstyle R}}{2})^{-\f{1}{2}} \  C',
\eeqn
and $g_+^{\scriptscriptstyle R} = g(\zeta_+^{\scriptscriptstyle R})$.
\subsection{Matching WKB and parabolic cylinder wave functions : Energy quantization equation for the lower energy level}
Now we are ready to match the WKB wave functions \eqref{WKBLB},\eqref{WKBBR} in regions $LB$ and $BR$,  respectively, with the parabolic cylinder wave functions \eqref{psi0} in those regions. By matching the wave functions in $LB$ we obtain relations between $C$, $C'$ and $\alpha_L$,
\beqn
\label{C1b}
C &=& (\f{\hbar \omega_L}{\pi m})^{\f{1}{4}} (\f{g_+^{\scriptscriptstyle L}}{2})^{\f{1}{2}}  \cos \pi \zeta_+^{\scriptscriptstyle L} \ \alpha_L,
\\ \label{C2b}
C' &=& \pi (\f{\hbar \omega_L}{\pi m})^{\f{1}{4}}  (\f{g_+^{\scriptscriptstyle L}}{2})^{-\f{1}{2}} e^{I_+}\Gamma^{-1}(-\zeta_+^{\scriptscriptstyle L})   \ \alpha_L. 
\eeqn
Matching the wave functions in \textit{BR} relate $C$, $C'$ to $\alpha_R$,
\beqn \label{C1a}
C &=& \pi (\f{\hbar \omega_R}{\pi m})^{\f{1}{4}}  (\f{g_+^{\scriptscriptstyle R}}{2})^{-\f{1}{2}} e^{I_+}\Gamma^{-1}(-\zeta_+^{\scriptscriptstyle R})   \ \alpha_R, 
\\ \label{C2a}
C' &=& (\f{\hbar \omega_R}{\pi m})^{\f{1}{4}} (\f{g_+^{\scriptscriptstyle R}}{2})^{\f{1}{2}}  \cos \pi \zeta_+^{\scriptscriptstyle R} \ \alpha_R.
\eeqn
In order to find the energy quantization equation we find the ratio $C/C'$ from Eqs. (\ref{C1b}-\ref{C2b}) and from Eqs.(\ref{C1a}-\ref{C2a}) and equate them. This gives us
\beq \label{pq}
\zeta_+^{\scriptscriptstyle L} \zeta_+^{\scriptscriptstyle R} = f(\zeta_+^{\scriptscriptstyle L}) f(\zeta_+^{\scriptscriptstyle R}) e^{-2I_+} 
\eeq
where
\beq
f(\zeta) = (2 \pi)^{-1} \cos \pi \zeta \ \Gamma(1-\zeta) \ g(\zeta).
\eeq
and we used the identity $t \Gamma(t) = \Gamma(1+t)$. Eq.\eqref{pq} is the fundamental equation of this section we were seeking. 
\subsection{Energy quantization equation for the upper energy level}
For the upper level in the doublet with energy $E_-$ one can similarly define 
\beqn
\zeta_-^{\scriptscriptstyle L} &\equiv& \f{E_-}{\hbar \omega_L} - \f{1}{2}, \\
\zeta_-^{\scriptscriptstyle R} &\equiv& \f{E_- - \tilde{\epsilon}}{\hbar \omega_R} - \f{1}{2}.
\eeqn
and do the previous procedure to obtain identically the energy equation 
\beq \label{pq+}
\zeta_-^{\scriptscriptstyle L} \zeta_-^{\scriptscriptstyle R} = f(\zeta_-^{\scriptscriptstyle L}) f(\zeta_-^{\scriptscriptstyle R}) \ e^{-2I_-} 
\eeq
where
\beq
I_- \equiv \int\limits_{a_-}^{b_-} \abs{p} dx.
\eeq

\section{Energy Splitting and Tunneling Amplitude to First order} \label{EnergySplitting}
Eqs.\eqref{pq},\eqref{pq+}, and in short
\beq \label{Energy}
\zeta_\pm^{\scriptscriptstyle L} \zeta_\pm^{\scriptscriptstyle R} = f(\zeta_\pm^{\scriptscriptstyle L}) f(\zeta_\pm^{\scriptscriptstyle R}) e^{-2I_\pm} ,
\eeq
 are transcendental equations. We can only solve them approximately. For small energy bias, $\tilde{\epsilon}/\hbar \omega \ll 1$, definitions of $\zeta_{\pm}^{\scriptscriptstyle L}$ and $\zeta_{\pm}^{\scriptscriptstyle R}$ imply that $\zeta_{\pm}^{\scriptscriptstyle L}, \zeta_{\pm}^{\scriptscriptstyle R} \ll 1$. For small values of $\zeta_{\pm}^{\scriptscriptstyle L}$ and $\zeta_{\pm}^{\scriptscriptstyle R}$,  $f(\zeta_\pm^{\scriptscriptstyle L}) f(\zeta_\pm^{\scriptscriptstyle R})$ is of order one hundredth. One can see this by expanding $f(\zeta)$ about zero and obtaining
\beq
f(\zeta) = \f{1}{\sqrt{4 e \pi}} (1 + k \ \zeta + \mathcal{O}(\zeta^2))
\eeq
where $k \simeq .11$ is defined in Eq. \eqref{k}. One then notes that for $f(\zeta) f(\zeta')$ the leading order term is $1/{4 e \pi} \simeq 0.02 \ $.  Now since $exp{(-2 I_{\pm})}$ is exponentially small in the WKB limit, the left hand side of Eq. \eqref{Energy} is also exponentially small. Therefore we can expand the right hand sides of Eqs.\eqref{pq},\eqref{pq+} to first order in $\zeta_{\pm}^{\scriptscriptstyle L}$ and $\zeta_{\pm}^{\scriptscriptstyle R}$ to find the energy levels in the ground state doublet, $E_{\pm}$,
\beq \label{pqm}
\zeta_{\pm}^{\scriptscriptstyle L} \zeta_{\pm}^{\scriptscriptstyle R} = \left[1 + k (\zeta_{\pm}^{\scriptscriptstyle L} + \zeta_{\pm}^{\scriptscriptstyle R}) + \mathcal{O}(\zeta_{\pm}^{{\scriptscriptstyle L}^2}, \zeta_{\pm}^{{\scriptscriptstyle R}^2})\right] \f{e^{-2 I_\pm}}{ 4 e \pi}.
\eeq
Now instead of engaging with $E_{\pm}$ which is very large compared to the tunnel splitting we define exponentially small quantities $\Delta E_{\pm}$ as follows and try to find them
\beqn \label{DeltaEpm}
\Delta E_{\pm} \equiv E_{\pm} - \bar{E}
\eeqn
where $\bar{E}$ is defined in Eq. \eqref{Ebar0}.
In terms of $\Delta E_{\pm}$ , one can write  
\beqn
\zeta_{\pm}^{\scriptscriptstyle L} &=& (\Delta E_{\pm} + \epsilon/2) / \hbar \omega_L, \\
\zeta_{\pm}^{\scriptscriptstyle R} &=& (\Delta E_{\pm} - \epsilon/2) / \hbar \omega_R
\eeqn
Please observe the appearance of $\epsilon$ instead of $\tilde{\epsilon}$. One can also expand $I_{\pm} (E_{\pm})$ around $\bar{E}$ to express both $\Delta{E_{\pm}}$ in terms of quantities defined at the mean energy $\bar{E}$,
\beq \label{I}
I_{\pm} = I(E_{\pm}) =  \bar{I} + \bar{I'} \Delta E_{\pm} + \mathcal{O} (\Delta E_{\pm}^2)
\eeq
where $\bar{I} = I(\bar{E})$ and $\bar{I'} = \f{\p I}{\p E} (\bar{E})$.\bibnote{The Taylor expansion is possible here because $\bar{E} \ne 0$. This is different from the case studied in Sec. III of Ref. \cite{Garg2000} where the expansion is taken around zero energy $E=0$.}

Now by using Eqs. (\ref{DeltaEpm}-\ref{I}), we can write Eq. (\ref{pqm}) as 
\beqn
\nn \f{\Delta E_{\pm}^2}{\hbar^2 \omega_R \omega_L} -  \f{(\epsilon/2)^2}{\hbar^2 \omega_R \omega_L} &=& \f{e^{-2 \bar{I}}}{4 \pi e} \left( 1 + \f{k \epsilon (\hbar \omega_R - \hbar \omega_L)}{2 \hbar^2 \omega_R \omega_L} \right) \\ \nn
&-& \f{e^{-2 \bar{I}}}{4 \pi e} \left(u \Delta E_{\pm} + \mathcal{O}(\Delta E_{\pm}^2) \right) \\ \label{DeltaE}
\eeqn
where 
\beq
u \equiv 2 \bar{I'} - \f{k \hbar (\omega_R + \omega_L)}{\hbar^2 \omega_R \omega_L}
\eeq
Since $e^{-2 \bar{I}}$ is exponentially small, we can reasonably neglect terms of order $e^{-2 \bar{I}}\Delta E_{\pm}^2$ in the right hand side of Eq. \eqref{DeltaE} while keeping the term of order $\Delta E_{\pm}^2$ in the left hand side of the equation. Eq. \eqref{DeltaE} then becomes a quadratic equation with two solutions as follows
\beqn
\Delta E_{\pm} = -b' \mp \sqrt{(\f{\epsilon}{2})^2 + (\f{\Delta}{2})^2 + b'^2 }
\eeqn
where
\beqn
b' &=& \f{\hbar^2 \omega_L \omega_R e^{-2 \bar{I}} u }{8 \pi e}, \\ \label{Delta2}
\Delta^2 &=& \f{\hbar^2 \omega_R \omega_L  e^{-2 \bar{I}}}{e \pi} \left(1 + k \f{\epsilon (\omega_R - \omega_L)}{2 \hbar \omega_R \omega_L} \right).
\eeqn
The level splitting $\Delta E$ can now be obtained,
\beqn
\nn \Delta E &=& E_- - E_+ = \Delta E_- - \Delta E_+ \\
&=& \sqrt{\epsilon^2 + \Delta^2 + (2 b')^2 }
\eeqn
The last term above, $(2 b')^2$, is of order $e^{- 4 \bar{I}}$ and can be neglected in favor of the second term, $\Delta^2$, which is of order $e^{-2 \bar{I}}$. This is irrespective of the value of $\epsilon$. Therefore, we obtain
\beq
\Delta E \cong \sqrt{\epsilon^2 + \Delta^2 }
\eeq
The second term above can be interpreted  as the square of tunneling matrix element. To take square root from right hand side of Eq. \eqref{Delta2} we note that $\epsilon/\hbar \omega_L \ll 1$ and $k,\f{\omega_L-\omega_R}{\omega_R} < 1$, so we can keep terms to first order in $\epsilon/\hbar \omega_L \ll 1$ and obtain
\beq \label{Delta-asym}
\Delta = \f{\hbar \sqrt{\omega_R \omega_L}}{\sqrt{e \pi}} (1 + \f{k}{4}\f{\epsilon}{\hbar \omega_L} \f{\omega_R - \omega_L}{\omega_R})e^{-\bar{I}} 
\eeq
which is the same as Eq. \eqref{Delta_asym} as promised. 

\section{Dependence of tunnel splitting on bias energy} \label{Dependence}
It has been believed \cite{Dekker87} that the dependence of $\Delta$ on $\epsilon$ is only through the quantity $\epsilon/V_0$ which is negligible  in the WKB limit. We are going to illustrate in this section that the dependence is also through the quantity $\epsilon/\hbar \omega$ which is much larger than $\epsilon/V_0$ and may not be neglected. This fact is rather clear from Eq. \eqref{Delta-asym} if $\omega_R \ne \omega_L$ in the unbiased double well potential when $\tilde{\epsilon} = 0$. That is to say if the potential in the absence of energy bias is not perfectly symmetric.  

The above argument is quite irrespective to the way the exponential factor $e^{-\bar{I}}$ in Eq. \eqref{Delta-asym} varies with $\epsilon$. What we wish to illustrate below, in addition, is that $e^{-\bar{I}}$ \textit{also} varies with  $\epsilon/\hbar \omega$ or $(\epsilon/\hbar \omega)^2$ as its largest correction. 

We are going to analyze below the Gamow factor $e^{-\bar{I}}$ analytically as much as possible and also illustrate numerical results for the dependence of the factor to $\epsilon/\hbar \omega$. However in order to build intuition and also give a counter example for the claim that correction to tunneling amplitude is of order $\epsilon/V_0$ let us begin by considering a simple example of a double oscillator potential.
\subsection{Example}
Consider a biased double oscillator potential 
\beq \label{VCounter}
V(x) =  \begin{cases} 
       \f{1}{2} m \omega_L^2 (x-x_L)^2  & x \le 0  \\
      \tilde{\epsilon} + \f{1}{2} m \omega_R^2 (x-x_R)^2 & x \ge 0 
   \end{cases} 
\eeq 
for appropriate values of $x_L < 0$, $x_R > 0$, $\omega_L$ and $\omega_R$. The height of the potential barrier is 
\beq
V_0 = \f{1}{2} m \omega_L^2 x_L^2 = \tilde{\epsilon} + \f{1}{2} m \omega_R^2 x_R^2
\eeq
We can freely choose $V_0$, $\omega_L$, $\omega_R$, $\tilde{\epsilon}$ and let the above constraint determine $x_L$ and $x_R$. To satisfy the WKB condition we just need to make sure that $V_0$ and $x_R-x_L$ are sufficiently large and $\omega_R$ and $\omega_L$ are not too large. Otherwise these quantities can be chosen freely.  The potential of Eq.(\ref{VCounter}) has a spike at the peak, at $x=0$, which violates the WKB condition $\f{m \hbar}{p^3} \f{dV}{dx} \ll 1$ \cite{Landau58}. However, one can smooth the potential near the spike such that the WKB condition is satisfied and the integrals of momentum stays almost intact. We continue with the potential of Eq. \eqref{VCounter} for its simplicity in calculations of the integrals and that we are only interested here in the mathematical properties of $e^{-\bar{I}}$. 

The integral 
\beq \label{I_bar}
\bar{I} = \f{1}{\hbar} \int\limits_{\bar{a}}^{\bar{b}} \abs{p} dx 
\eeq
can be divided into two parts 
\beq \label{Ibar_ILIR}
\bar{I} = \bar{I}_L + \bar{I}_R
\eeq
where
\beq \label{IL_and_IR}
\bar{I}_L = \f{1}{\hbar} \int\limits_{\bar{a}}^{x_m} \abs{p} dx \ , \quad \bar{I}_R = \f{1}{\hbar} \int\limits_{x_m}^{\bar{b}} \abs{p} dx 
\eeq
and where $x_m$ is the coordinate of the maximum potential. This is the strategy we shall use in the next subsection too for analytic study of the general case.  For the potential of Eq.(\ref{VCounter}), $\bar{I}_L$ can be easily calculated. The result in terms of the energy quantities is
\beq
\bar{I}_L = \f{V_0}{\hbar \omega_L} \left( \sqrt{1 - \lambda_L } - \lambda_L \log \f{\sqrt{1 - \lambda_L} + 1}{\lambda_L} \right)
\eeq 
where $\lambda_L = \bar{E}/V_0$. Similarly for $\bar{I}_R$ one obtains, 
\beq
\bar{I}_R = \f{V_0 - \tilde{\epsilon}}{\hbar \omega_R} \left( \sqrt{1 - \lambda_R } - \lambda_R \log \f{\sqrt{1 - \lambda_R} + 1}{\lambda_R} \right)
\eeq 
where $\lambda_R = (\bar{E}- \tilde{\epsilon})/(V_0-\tilde{\epsilon})$. Since $\lambda_L,\lambda_R \ll 1$ we can expand the above expression in terms of $\lambda_L,\lambda_R$,
\begin{align}
\label{IL} \bar{I}_L &=& \f{V_0}{\hbar \omega_L} \left( 1 + \f{\lambda_L}{2} \{\log(\f{\lambda_L}{2}) - \f{1}{2}\} + \mathcal{O}(\lambda_L^2) \right) \\
\label{IR} \bar{I}_R &=& \f{V_0 - \tilde{\epsilon}}{\hbar \omega_R} \left( 1 + \f{\lambda_R}{2} \{\log(\f{\lambda_R}{2}) - \f{1}{2}\} + \mathcal{O}(\lambda_R^2) \right) 
\end{align}
Now we note that 
\beqn
\f{V_0}{\hbar \omega_L} \lambda_L &=& \f{\bar{E}}{\hbar \omega_L}, \\
\f{V_0 -\tilde{\epsilon}}{\hbar \omega_R} \lambda_R &=& \f{\bar{E} - \tilde{\epsilon}}{\hbar \omega_R}.
\eeqn
This shows that the largest correction is of $\tilde{\epsilon}/\hbar \omega$. To see it more clearly we combine Eqs. (\ref{IL}-\ref{IR}) to obtain $\bar{I}$ to leading orders
\beqn
\nn \bar{I} &=& \f{V_0}{\hbar \omega_L} + \f{V_0 - \tilde{\epsilon}}{\hbar \omega_R} + (\f{1}{4}+ \f{\epsilon}{4 \hbar \omega_L}) \{\log(\f{\lambda_L}{2}) - \f{1}{2}\} \\ \label{Ibar}
&+& (\f{1}{4}-  \f{\epsilon}{4 \hbar \omega_R}) \{\log(\f{\lambda_R}{2}) - \f{1}{2}\} + \mathcal{O}(\f{\bar{E}}{V_0}, \f{\tilde{\epsilon}}{V_0})
\eeqn
Here we used the identities 
\beqn
\label{Ebar} \bar{E} &=& \f{\hbar \omega_L}{2} + \f{\epsilon}{2} \\
\label{Ebar-e}\bar{E} - \tilde{\epsilon} &=& \f{\hbar \omega_R}{2} - \f{\epsilon}{2}
\eeqn
which can be obtained from the definitions of $\bar{E}$ and $\epsilon$ in Eqs. (\ref{Ebar0}), (\ref{epsilon}). To zeroth order in $\tilde{\epsilon}/V_0$, we have $\lambda_R = \lambda_L \equiv \lambda$. Therefore Eq. (\ref{Ibar}) becomes
\begin{align}
\nn \bar{I} &= \f{V_0}{\hbar \omega_L} + \f{V_0}{\hbar \omega_R} - \f{\tilde{\epsilon}}{\hbar \omega_R} \\ 
&+ (\f{1}{2}+ \f{\epsilon}{4 \hbar \omega_L} \f{\Delta \omega}{\omega_R}) \{\log(\f{\lambda}{2}) - \f{1}{2}\} + \mathcal{O}(\f{\bar{E}}{V_0}, \f{\tilde{\epsilon}}{V_0}) 
\end{align}
One now observes that the leading order correction, due to the bias, comes from the third term in the right hand side of the above equation, i.e. from $-\tilde{\epsilon}/\hbar \omega_R$.  In the case that $\omega_L \ne \omega_R$ the correction from the fourth term, $\f{\epsilon}{4 \hbar \omega_L} \f{\Delta \omega}{\omega_R} \log(\f{\lambda}{2})$, is also important. In fact this can be the dominant correction if $\lambda$ is suitably small. We did not keep terms of order $\mathcal{O}(\f{\bar{E}}{V_0})$ above. One could keep them, but that would not alter the conclusion if one neglects terms of order $\mathcal{O}(\tilde{\epsilon}/V_0)$ and $\mathcal{O}(\epsilon/V_0)$. This completes our counter example for the statement which had expressed that the corrections are of order $\tilde{\epsilon}/V_0$. 

\subsection{General Case}
For a general double well potential we again divide $\bar{I}$ into $\bar{I}_L$ and $\bar{I}_R$ as in Eqs. (\ref{I_bar}-\ref{IL_and_IR}). Then we use the results of Sec. III of Ref. \cite{Garg2000} which deals with a similar integral (Ref. \cite{Garg2000} solves the problem of \emph{symmetric} potential. However some of integral calculus done there can be used here if one does division (\ref{I_bar}-\ref{IL_and_IR}) for the action). We combine Eqs. (3.5), (3.10), and (3.11) of \cite{Garg2000} for $\bar{I}_L = I_L (\bar{E})$ to obtain 
\beqn
\nn I_L (\bar{E}) &=& I_L (0)  - \f{m \omega_L}{2 \hbar} (\bar{a}-x_L)^2 \log \f{2 (x_m - x_L)}{\bar{a}-x_L} \\ \nn &-& \f{m \omega_L}{2 \hbar} (\bar{a}-x_L)^2 (A_L + \f{1}{2}) \\ \label{IL(Ebar)} &+& \mathcal{O}((\bar{a}-x_L)^3) 
\eeqn
where 
\beqn
I_L (0) &=& \f{1}{\hbar} \int\limits_{x_L}^{x_m} \sqrt{2 m V(x)} \\
A_L &=& \int\limits_{x_L}^{x_m} \{\f{m \omega_L}{\sqrt{2 m V(x)}} - \f{1}{x-x_L} \} dx
\eeqn
$A_L$ is of order unity. For example for a symmetric quartic double well potential $A_L = \log 2$ (see e.g. Sec. V of Ref. \cite{Garg2000}). $I_L(0)$ would be half of the action if the potential were symmetric. In deriving Eq.(\ref{IL(Ebar)}) one approximates the potential with a parabola near the minimum $x_L$ all the way to the turning point $\bar{a}$. We can use this approximation to write Eq. (\ref{IL(Ebar)}) in terms of $\bar{E}$ by noting that $\bar{E} \simeq \f{1}{2} m \omega_L^2 (\bar{a} - x_L)^2$: 
\beqn
\nn I_L (\bar{E}) &\simeq& I_L (0) - \f{\bar{E}}{2 \hbar \omega_L} \log \f{2 (x_m - x_L)}{\sqrt{2 \bar{E} / m \omega_L^2}} \\ \label{IL_Ebar} &-& \f{\bar{E}}{2 \hbar \omega_L}  (A_L + \f{1}{2}) + \cdots 
\eeqn
In virtue of Eq. \eqref{Ebar} one can observe that $I_L(\bar{E})$ in the above equation has corrections of order $\epsilon/\hbar \omega_L$. To expand $\bar{I}_R= I_R(\bar{E})$ we use the same strategy as in Sec. III of Ref. \cite{Garg2000}. The only change that is required is to shift the zero point of potential up by the amount $\tilde{\epsilon}$. Then all the arguments trivially follow and we obtain 
\beqn
\nn I_R (\bar{E}) &\simeq& I_R (0) - \f{\bar{E} - \tilde{\epsilon}}{2 \hbar \omega_R} \log \f{2 (x_R - x_m)}{\sqrt{2 (\bar{E}- \tilde{\epsilon}) / m \omega_R^2}} \\ \label{IR_Ebar} &-& \f{\bar{E}-\tilde{\epsilon}}{2 \hbar \omega_R}  (A_R + \f{1}{2}) + \cdots 
\eeqn
where
\beqn
I_R (0) &=& \f{1}{\hbar} \int\limits_{x_m}^{x_R} \sqrt{2 m (V(x) - \tilde{\epsilon}) } \\
A_R &=& \int\limits_{x_m}^{x_R} \{\f{m \omega_R}{\sqrt{2 m (V(x)- \tilde{\epsilon}) }} - \f{1}{x_R-x} \} dx
\eeqn
Using Eq. \eqref{Ebar-e} one can see that $I_R(\bar{E})$ also has corrections of order $\epsilon/\hbar \omega_R$. So in general first order correction of order $\te/\hbar \omega$ or $\epsilon/\hbar \omega$ appears in the tunneling amplitude both from the prefactor and the Gamow factor, 
\beq \label{Delta-asym-l}
\Delta (\te) = \f{\hbar \sqrt{\omega_R \omega_L}}{\sqrt{e \pi}} (1 + \f{k}{4}\f{\epsilon}{\hbar \omega_L} \f{\omega_R - \omega_L}{\omega_R})e^{-[\bar{I}_L(\te) + \bar{I}_R(\te)]} 
\eeq
In some circumstances all the first order corrections cancel. This happens if there is unitary transformation between the Hamiltonians of the same potential with positive and negative bias of the same magnitude as we discuss elsewhere along with the applications of first order correction in tunnel splitting. In general, however, one might expect to get such first order corrections in the tunneling amplitude. 

\section{Conclusion}

In conclusion, we did a WKB calculation in this paper to find the tunnel splitting in one dimensional asymmetric potentials. We found that the tunnel splitting can in general have first order dependence to the bias energy. We showed that the dependence is of order $\te/\hbar \omega$ which is greater  than $\te/V_0$ which was previously thought. 

\bibliographystyle{unsrt}
\bibliography{IOPEXPORT_BIB} 

\end{document}